# Space weathering simulations through controlled growth of iron nanoparticles on olivine

T. Kohout[1,2] (kohout@gli.cas.cz), J. Čuda[3], J. Filip[3], D. Britt[4], T. Bradley[4], J. Tuček[3], R. Skála[1], G. Kletetschka[1,5], J. Kašlík[3], O. Malina[3], K. Šišková[3], and R. Zbořil[3]

1. Institute of Geology, Academy of Sciences of the Czech Republic, Prague, Czech Republic

2. Department of Physics, University of Helsinki, Finland

3. Regional Centre of Advanced Technologies and Materials, Faculty of Science, Departments of Experimental Physics and Physical Chemistry, Palacký University, Olomouc, Czech Republic

4. Department of Physics, University of Central Florida, Orlando, USA

5. Faculty of Science, Charles University in Prague, Czech Republic

## Abstract

Airless planetary bodies are directly exposed to space weathering. The main spectral effects of space weathering are darkening, reduction in intensity of silicate mineral absorption bands, and an increase in the spectral slope towards longer wavelengths (reddening). Production of nanophase metallic iron ($npFe^0$) during space weathering plays major role in these spectral changes. A laboratory procedure for the controlled production of $npFe^0$ in silicate mineral powders has been developed. The method is based on a two-step thermal treatment of low-iron olivine, first in ambient air and then in hydrogen atmosphere. Through this process, a series of olivine powder samples was prepared with varying amounts of $npFe^0$ in the 7-20 nm size range. A logarithmic trend is observed between amount of $npFe^0$ and darkening, reduction of 1 μm olivine absorption band, reddening, and 1 μm band width. Olivine with a population of physically larger $npFe^0$ particles follows spectral trends similar to other samples, except for the reddening trend. This is interpreted as the larger, ~40-50 nm sized, $npFe^0$ particles do not contribute to the spectral slope change as efficiently as the smaller $npFe^0$ fraction. A linear trend is observed between the amount of $npFe^0$ and 1 µm band center position, most likely caused by $Fe^{2+}$ disassociation from olivine structure into $npFe^0$ particles.

## Introduction

Airless planetary bodies are directly exposed to the space environment and thus to space weathering. Space weathering is caused by a combination of solar wind and solar radiation, micrometeorite bombardment, and cosmic radiation that can alter the physical, chemical, and crystallographic properties of airless regoliths. Among other effects, space weathering causes changes in the visible, IR and UV spectra of exposed surface material making it difficult to compare weathered asteroidal and lunar spectra to those of unweathered silicate minerals, meteorites, and lunar samples.

### Lunar type space weathering

Lunar samples delivered to Earth by the Apollo program during the 1960's and 1970's significantly contributed to our understanding of space weathering. It was the first time that direct laboratory studies were possible on a material exposed for long periods to the space environment and, therefore, modified by space weathering. Numerous early studies (e.g. Adams and Jones, 1970, McCord and Johnson, 1970, Adams and McCord, 1971, 1973, McCord and Adams, 1973) found that, compared to artificially crushed fresh Apollo rock samples, spectra of lunar soils are darker, have reduced intensity of silicate mineral absorption bands, and their spectral slope increases towards longer wavelengths (reddening).

Reviews by Hapke (2001) or Chapman (2004) provide an understanding of the mechanisms involved in space weathering. Earlier lunar soil studies indicated that space weathering may cause an accumulation of dark agglutinitic glass formed by surface melting of regolith by micrometeorite impacts (Adams and McCord, 1971). Agglutinitic glass is an important optical agent influencing reflectance spectra of lunar regolith. However, its presence could not explain all of the spectral discrepancies in lunar soils and planetary surfaces, especially the spectral reddening (Pieters et al., 1993). Hapke et al. (1975) suggested that optical effects of space weathering may be caused by metallic submicroscopic iron particles (SMFe), also referred to as nanophase iron ($npFe^0$) particles, within vapor deposition rims on lunar soil grains. This $npFe^0$ is believed to originate from disassociation of Fe from Fe-bearing minerals through solar wind sputtering and micro-impact-generated vapor (e.g. Keller and McKay, 1993; 1997, Hapke, 2001), and its concentration is highest in the fine fraction of lunar soils (Pieters et al, 1993; Taylor et al., 2001). Support for this theory came later from transmission electron microscopy (TEM). TEM observations showed the presence of such $npFe^0$ particles in lunar soil grain coatings (e.g. Keller and McKay, 1997; Hapke, 2001). The typical size range of $npFe^0$ particles is ~1-10 nm in vapor deposition rims and ~10-100 nm in agglutinates (e.g. Keller and Clemett, 2001).

*S-type asteroid space weathering*

As in the case of lunar rocks, spectral discrepancies are also observed between meteorites and asteroids (e.g. Hapke, 2001 and Chapman, 2004). Silicate rich S-complex asteroids are represented by absorption bands characteristic of olivine (at wavelengths around 1 μm) and pyroxene (at wavelengths around 1 and 2 μm). Spectra of ordinary chondrite meteorites contain similar absorption bands, but, compared to asteroids, the bands in meteorites are usually more intense and the overall spectral slope is flatter. Numerous laboratory simulations (involving laser irradiation and ion bombardment, e.g. Brunetto et al., 2005) of ordinary chondrites and detailed spectral observations of asteroids and meteorites (e.g. Marchi et al., 2005; Lazzarin et al., 2006) show that space weathering effects on chondritic materials and S-complex asteroids display similarities to lunar soils where solar wind ion bombardment induces atom displacements and micrometeorite bombardment forms $npFe^0$ particles. Direct evidence of $npFe^0$ particle formation on the surface of S-type asteroids was recently reported by Noguchi et al. (2011) in a study of surface coatings of regolith grains obtained from asteroid 25143 Itokawa by the Hayabusa sample return mission. However, compared to lunar weathering, the mechanism of asteroidal space weathering is more complex and its various effects (e.g. absorption band weakening, darkening, and reddening) do occur in spectra, with a highly varying intensity, even on the same asteroids (e.g. Chapman, 2004; Hiroi et al., 2006; Gaffey, 2010).

Observations of young asteroid families shows that space weathering occurs relatively rapidly, within $10^6$ years after the breakup of the parent body (Vernazza et al., 2009). Similar to lunar type weathering, space weathering on asteroids affects a thin surface regolith layer only and this layer can be disturbed by geological processes such as landslides, as observed on Eros by Clark et al. (2001), or by tidal forces during close planetary encounters (Binzel et al., 2010).

*Space weathering laboratory simulations*

Laboratory production of $npFe^0$ particles similar to those observed in space-weathered lunar soils was achieved by direct mineral or glass reduction (Allen et al., 1993), short duration laser irradiation (e.g. Sasaki et al., 2001; 2002; 2003; Brunetto et al., 2005; Lazzarin et al., 2006; Markley et al., 2013), ion bombardment (e.g. Marchi et al.; 2005, Lazzarin et al.; 2006), direct mineral / glass synthesis (Liu et al, 2007), or microwave irradiation (Tang et al., 2012) on various silicate minerals, meteorites or lunar rocks. Since then, laser irradiation together with ion bombardment has become a common laboratory tool for space weathering simulation, capable of reproducing most spectral changes. However, these methods do not provide sufficient control over npFe$^0$ particle size and concentration and, thus, do not enable quantitative space weathering simulations.

Only a handful of studies were devoted to control the size and quantity of npFe$^0$ and to verify its effect on spectral properties. For example, Allen et al. (1996) produced 6 nm sized npFe$^0$ particles on a silica gel substrate with 6 nm pores through impregnating the pores with ferric nitrate solution and subsequent reduction of iron in a hydrogen atmosphere. Noble et al. (2007) enhanced this technique with silica gel substrates featuring pore sizes of 2.3, 6, 25 and 50 nm resulting in a better control of npFe$^0$ size and concentration. The changes in optical properties of such treated samples resembled the space weathering effects observed in lunar soils and it was possible to evaluate the influence of npFe$^0$ particle size on spectral red slope.

In this work we present a new method of controlled npFe$^0$ production on surface of olivine grains. We also evaluate the influence of npFe$^0$ concentration and particle-size on olivine spectral darkening and reddening, as well as intensity reduction, shape, and position of the olivine 1 μm absorption band.

Materials and methods

*Samples*

Natural olivine from Åheim, Norway was used in this study. A pale green, polycrystalline olivine sample was thin-sectioned, and analyzed using a CAMECA SX-100 electron microprobe (Table 1). Another part of the sample was crushed and inclusion-free fragments were hand-picked using an optical microscope. Subsequently, the olivine grains were pulverized in an agate mortar and sieved to obtain fine (10-80 µm) powder. Part of the powder was preserved for reference measurements while the rest was thermally treated in a two-step heating method to achieve the formation of $npFe^0$ on the surface of olivine grains.

*$npFe^0$ production and characterization*

A two-step thermal treatment method was developed to produce $npFe^0$ on the surface of olivine grains. Furthermore, $npFe^0$ abundance and particle size control was achieved through variations in temperature and duration of the treatment steps.

In the first step, the olivine powder was heated in air (under oxidizing condition) using a Linn LM 112.07 muffle furnace. Various temperatures and heating durations were tested (Table 2) in order to induce partial oxidation of the iron ions that were liberated from the olivine structure, primarily those close to the surface of the olivine grains. The temperatures were selected to be low enough to preserve the olivine structure undamaged (e.g. Barcova et al., 2003; Michel et al., 2013). The oxidation of $Fe^{2+}$ ions leads to a charge imbalance in the olivine structure causing diffusion of $Fe^{3+}$ atoms to the surface where they precipitate in the form of $Fe^{3+}$ oxide nanoparticles and create vacancies in the olivine crystal structure (Zboril et al., 2003; Barcova et al., 2003).

In the second step, the $Fe^{3+}$ oxide nanoparticles were reduced into $npFe^0$ in a hydrogen atmosphere (using a steady hydrogen flow at 500°C for 1 hour with an Anton Paar XRK900 reactor). The reduction process was monitored in-situ using X-ray powder diffraction (XRD). The freshly produced $npFe^0$ particles were surface-passivated for 30 minutes in 30°C gas mixture flow of $N_2$ with 2% $O_2$ to minimize oxidation upon contact with ambient air.

The XRD was performed using a PANalytical X´Pert PRO MPD diffractometer (iron-filtered $CoK_\alpha$ radiation: $\lambda$ = 0.178901 nm, 40 kV and 30 mA) in the Bragg-Brentano geometry and equipped with an X´Celerator detector, that has programmable divergence and diffracted beam anti-scatter slits.

The samples were placed into a shallow cavity sample holder (made of Macor) and repeatedly-scanned in the 2θ range of 5–120° (resolution of 0.017° 2θ) at specific temperatures. SRM640 (Si) and SRM660 ($LaB_6$) commercial standards from NIST (National Institute of Standards and Technology) were used for evaluation of the line positions and instrumental line broadening, respectively. Identification of crystalline phases was performed using the HighScore Plus (PANalytical) software in conjunction with the PDF-4+ database.

The size and concentration of $npFe^0$ was determined using a combination of transmission electron microscopy (TEM), frequency dependence of magnetic susceptibility, and magnetic hysteresis measurements, Composition of the $npFe^0$ was verified using scanning transmission electron microscopy equipped with energy-dispersive X-ray spectroscopy (STEM-EDX).

The TEM observations and STEM-EDX analysis was done using a Teknai F30 TEM/STEM with an EDAX Si/Li EDX spectrometer. The TEM observations of $npFe^0$ particles were done on sharp, thin edges of olivine powder grains placed on a copper grid holder.

The frequency dependence of magnetic susceptibility was tested using a ZH instruments SM-100/105 susceptibility meters at 16, 32, 64, 128 and 256 kHz frequency steps and 160 A/m RMS field intensity.

The magnetic hysteresis measurements were accomplished using a Princeton Measurements Micromag Model 3900 VSM (Vibrating Sample Magnetometer). All magnetizations were normalized by sample. First, a hysteresis loop of the olivine precursor was measured. This loop was subsequently used for background subtraction (to subtract paramagnetic/diamagnetic contribution of olivine and sample holder) of all thermally-treated olivine sample hysteresis loops. After background subtraction, any residual paramagnetic/diamagnetic slope was removed. Resulting hysteresis loops correspond to $npFe^0$ produced in our samples. The concentration of

npFe$^0$ can be estimated by comparison of the sample saturation magnetization to that of a pure metallic iron (218 Am$^2$/kg, Dunlop and Özdemir, 2001, p. 51).

For room-temperature $^{57}$Fe Mössbauer measurements, a Mössbauer spectrometer with a $^{57}$Co(Rh) source of γ-rays was used. The values of the derived hyperfine Mössbauer parameters are attributed to metallic iron (α-Fe) at room temperature. Mössbauer spectra were fitted by means of the Lorentzian line shapes using the least squares method featured in the MossWinn analysis program. To obtain a qualitatively resolved Mössbauer spectrum for low npFe$^0$ concentration, a long measurement time was applied (22 days).

The spectral measurements of fresh and modified olivine samples were done in visible – near infrared (VIS-NIR) range of 350-2400 nm using an Analytical Spectral Devices FieldSpec Pro spectrometer calibrated with a Labsphere SRS-99-020 Spectralon white standard. The reflectance is determined at 550 nm. The spectral slope is calculated as the difference in normalized reflectance at 1689 nm and 630 nm divided by 1059 nm. The normalized reflectance was calculated as reflectance divided by reflectance at 550 nm. The 1 μm absorption band depth is calculated as the normalized average reflectance at 630 and 1689 nm minus the band minimum. The 1 μm absorption band minimum position was manually read from the data files. To further study shift of the band minimum in our samples the modified Gaussian model (MGM, Sunshine et al., 1999, MATLAB code available on RELAB web page http://www.planetary.brown.edu/mgm/index.html) was applied to the spectral data.

Results

The combination of the two successive thermal treatment steps resulted in a formation of npFe$^0$ (α-Fe, bcc structure) on the olivine grains with a well-controlled particle size and distribution over the surface. The characteristics of such npFe$^0$ particles is comparable to the npFe$^0$ observed in space weathered extraterrestrial materials. Table 3 summarizes spectral properties of the thermally-treated olivine samples and npFe$^0$ concentration estimate derived from magnetic hysteresis measurements. There is a quasi-exponential trend in the npFe$^0$ concentration with increasing temperature of the first heating step (Fig. 1). Longer duration of the first heating step also increases the npFe$^0$ concentration. Due to the quasi-exponential trend between npFe$^0$ concentration and first step heating temperature, the samples heated to temperatures in excess of 750°C become very rich in npFe$^0$ and their reflectance spectra become very low. Thus, in order to reduce the amount of npFe$^0$ in the 850°C sample to the level of other samples, the duration of the first heating step was reduced to 30 seconds.

The size of the npFe$^0$ particles in all samples as seen by TEM (Fig. 2) is in ~5-20 nm range with the majority (~90%) being in the 7-15 nm range. An exception is the sample heated to 850°C (850s30) where an additional population of larger ~40-50 nm npFe$^0$ particles can be observed (Fig. 3). The larger particles most likely formed by sintering of smaller particles into clusters.

The above mentioned npFe$^0$ particle range is supported by a frequency dependence of the magnetic susceptibility data which occurs in particles that are in a superparamagnetic (SP) state that requires iron particles smaller than ~8 nm (Kneller and Luborsky, 1963; Dunlop and Özdemir, 2001, p. 131). No detectable systematic frequency dependence was observed in any sample, indicating that majority of the npFe$^0$ particles were larger than the ~8 nm threshold.

One single npFe$^0$ particle was investigated by high-resolution TEM in order to determine the lattice spacing on the single nanoparticle. Lattice fringes of 0.19 ±0.05 nm were observed (Fig. 4) which is consistent with the spacing of (110) of α-Fe $d_{110}$ = 0.203 nm (Noguchi et al., 2011). The surface passivation of npFe$^0$ on the olivine limited the formation of iron oxides to a thin shell on the metallic nanoparticle core (cf. Filip et al., 2007 and Siskova et al., 2012). This is well documented on STEM-EDX data as little oxygen (mostly originating from olivine background) is detected in individual npFe$^0$ particles (Fig. 5).

Additionally, the modified olivine samples were studied by means of zero-field $^{57}$Fe Mössbauer spectroscopy at room temperature. Only paramagnetic doublet (isomer shift $\delta$ = 1.12 mm/s and quadrupole split $\varepsilon_Q$ = 2.96 mm/s) and the singlet ($\delta$ = 0.53 mm/s), both originating from Fe$^{2+}$ in the two olivine octahedral sites with different point symmetries, are visible in the spectra. The ferromagnetic sextet component corresponding to ferromagnetic metallic iron is not observed due to npFe$^0$ amount being below Mössbauer spectroscopy resolution.

The reflectance spectra of fresh and modified olivine samples were measured in 350-2400 nm (VIS-NIR) range. A progressive trend in the reflectance reduction (darkening), in 1 μm absorption band depth reduction, slope change (reddening), shift in 1 μm band center position, and 1 μm band width at half depth is observed with the increasing npFe$^0$ concentration (Fig. 6).

Discussion

The size range of the npFe$^0$ (~5-20 nm) produced in olivine powders by our method is slightly larger than that found in vapor deposition rims (~3 nm), but considerably smaller than the nanophase iron found in agglutinates (e.g. Pieters et al., 2000; Hapke, 2001, Keller and Clemett, 2001). The presence of the npFe$^0$ is also seen in the strong alteration of the VIS-NIR spectra. Compared to spectra of the fresh olivine, samples with artificially produced npFe$^0$ show spectral changes similar to these seen in naturally space-weathered lunar soils and asteroid surfaces (e.g. Hapke, 2001; Chapman, 2004). This gives us a confidence that despite slight differences between our samples and natural space weathering products (npFe$^0$ particles on mineral grain surfaces vs. within coating rims, possible presence of a thin oxide shell on our npFe$^0$ particles) our laboratory simulations are good proxy for natural space weathering and closely resemble optical effects of the fine nanophase iron fraction including slope change (Noble et al., 2007). The changes in olivine spectral parameters are quantified in Table 3. Because the $npFe^0$ particle size is kept constant (with exception of 850s30 sample), the spectral changes can be studied as a function of increasing $npFe^0$ concentration. The results indicate that there is a linear trend between amount of $npFe^0$ and the 1 µm band center position (Fig. 7) and a logarithmic trend between the amount of $npFe^0$ and the reflectance (Fig. 8), 1 µm absorption band depth (Fig. 8), spectral slope (Fig. 9) and the 1 µm band width at half depth (Fig. 10).

The linear trend between amount of $npFe^0$ and the 1 µm band center position is in contrast with results of previous studies (Hiroi and Sasaki, 2001; Sasaki et al., 2002; 2003; Brunetto et al,. 2005) where no significant changes of the band position were observed during simulated space weathering. However, compared to these studies, our samples cover a broader range of $npFe^0$ amount and related spectral changes. It is apparent that for samples with less $npFe^0$ (below 0.01 wt% and spectral changes of a similar magnitude as in above-mentioned studies) there is no significant change in band position. However, with higher $npFe^0$ the amount of the shift in 1 µm band center became more apparent. The mechanism behind the band shift may be $Fe^{2+}$ disassociation from olivine structure into $npFe^0$ particles. This is also supported by the linear correlation between 1 µm band center and amount of $npFe^0$. The possible occurrence of this process in extremely weathered natural planetary surfaces is uncertain and requires further studies. In general, asteroidal regoliths do not get as mature as lunar regolith and olivine is minor constituent of the lunar regolith.

The band center shift was further studied using the MGM model. As the olivine 1 µm absorption band is a superposition of three individual bands related to the different positions of $Fe^{2+}$ ions in the olivine crystal structure (Burns, 1970), three Gaussian bands were used in the MGM model. The results are summarized in table 4 and figure 11. MGM method gives reliable results for samples with small to moderate $npFe^0$ amounts up to sample 650h1 with 0.023 wt% metallic Fe. (With higher $npFe^0$ content the lower 1 μm band intensity to noise ratio and significant red slope causes unreliable Gaussian modeling of olivine samples.) The MGM results indicate that the shift in 1 μm band is primarily driven by changes in the third Gaussian band centered around 1200 nm (Fig. 11) which is caused (together with the ~840 nm band) by the $Fe^{2+}$ in $M(2)$ position (e.g. Burns, 1970). The MGM model figures are available in on-line supplementary material (Figs. S1-S7).

The observed logarithmic trend between slope change and amount of $npFe^0$ provides further insight into natural space weathering. A logarithmic trend between spectral slope change and the space weathering duration has been observed by Nesvorný et al. (2005) and Vernazza et al. (2009) for S-type asteroid families. In combination with our observation of a logarithmic trend between amount of $npFe^0$ and reflectance, 1 μm absorption band depth, and spectral slope, we have demonstrated four additional characteristics of space weathering: A similar logarithmic trend is valid between space weathering duration and (1) darkening, (2) reduction of 1 μm olivine absorption band, and (3) 1 μm band width at half depth. However, (4) the amount of $npFe^0$ increases linearly with the duration of the space weathering.

The 850s30 sample contains additional population of ~40-50 nm $npFe^0$ particles which accounts for its VIS-NIR spectral properties. The spectrum of this sample follows the trends observed for other samples. However, sample 850s30 does not fully follow the increasing red slope trend (Fig. 9). This is most likely due to the fact that part of its $npFe^0$ is in a form of larger particles, which does not contribute to the red slope. This sample is analogous to lunar soils containing both small $npFe^0$ fraction in vapor deposition rims as well as larger $npFe^0$ particles in agglutinates. The relatively reduced magnitude of the slope change in this sample is in agreement with Noble et al. (2007) and Lucey and Noble (2008) who observed insignificant slope change for $npFe^0$ particles larger than ~50 nm.

Conclusions

The two-step thermal treatment method allows for controlled growth of iron nanoparticles on the surfaces of olivine powder grains. This enables quantitative investigations of the role of npFe$^0$ in space weathering and related changes in reflectance spectra. Compared to fresh olivine, our olivine samples with artificially introduced ~5-20 nm sized npFe$^0$ particles exhibit the spectral characteristics of lunar-type space weathering. From a quantitative point of view, a linear trend is observed between the amount of npFe$^0$ and 1 μm band center position. This trend is more pronounced for samples with npFe$^0$ amounts in excess of 0.015 wt%. The mechanism behind the band shift may be Fe$^{2+}$ disassociation from olivine structure into npFe$^0$ particles.

A logarithmic trend is observed between amount of npFe$^0$ and darkening, reduction of 1 μm olivine absorption band, reddening, and 1 μm band width at half depth. Observations of asteroid families show a logarithmic weathering trend between slope change and duration. Our results reveal four additional characteristics of space weathering: The logarithmic trend with space weathering duration is also valid for (1) darkening, (2) reduction of 1 μm olivine absorption band, and (3) 1 μm band width at half depth, while (4) the amount of npFe$^0$ increases linearly with duration.

The olivine sample with an additional population of larger npFe$^0$ particles follows similar spectral trends as other samples, except for the reddening trend. This is interpreted as the larger, (~40-50 nm sized), npFe$^0$ particles do not contribute to the slope change as efficiently as the smaller npFe$^0$ fraction.

Acknowledgements

The work was supported by Ministry of Education, Youth and Sports of the Czech Republic (grant no. LH12079, LK21303, MSM0021620855), Academy of Finland (grant no. 257487), Czech Science Foundation (grant no. GACR P108/11/1350) and Palacký University, Olomouc, Czech Republic (grant no. PrF_2013_014). We also thank for the support by the Operational Program Research and Development for Innovations – European Regional Development Fund (CZ.1.05/2.1.00/03.0058) and Operational Program Education for Competitiveness – European Social Fund (CZ.1.07/2.3.00/20.0017, CZ.1.07/2.3.00/20.0170, CZ.1.07/2.3.00/20.0155, and CZ.1.07/2.3.00/20.0056) of the Ministry of Education, Youth and Sports of the Czech Republic. This work (publication SSERVI-2014-080) was directly supported by NASA's Solar System Exploration Research Virtual Institute cooperative agreement notice NNA14AB05A.

Table 1. Chemical analyses of olivine used in this study.

| | 1 | 2 | 3 | 4 | 5 | 6 | 7 | 8 | 9 | 10 | 11 | 12 | 13 | 14 | 15 | 16 | 17 | 18 | 19 | 20 | mean | s.d. |
|---|---|---|---|---|---|---|---|---|---|---|---|---|---|---|---|---|---|---|---|---|---|---|
| Analyses in wt% | | | | | | | | | | | | | | | | | | | | | | |
| SiO$_2$ | 40.88 | 41.25 | 41.75 | 41.33 | 41.23 | 41.26 | 41.38 | 41.44 | 41.19 | 41.18 | 41.09 | 41.77 | 41.52 | 41.33 | 41.64 | 41.47 | 41.58 | 41.41 | 41.94 | 41.95 | 41.43 | 0.28 |
| Al$_2$O$_3$ | 0.04 | b.d.l. | b.d.l. | b.d.l. | b.d.l. | b.d.l. | 0.04 | b.d.l. | b.d.l. | b.d.l. | b.d.l. | b.d.l. | b.d.l. | b.d.l. | b.d.l. | b.d.l. | b.d.l. | b.d.l. | b.d.l. | b.d.l. | 0.04 | 0.00 |
| FeO$^T$ | 6.69 | 6.67 | 6.58 | 6.72 | 6.67 | 6.80 | 6.63 | 6.69 | 6.72 | 6.74 | 6.67 | 6.64 | 6.71 | 6.75 | 6.63 | 6.68 | 6.67 | 6.61 | 6.63 | 6.68 | 6.68 | 0.05 |
| NiO | 0.39 | 0.37 | 0.41 | 0.39 | 0.36 | 0.38 | 0.38 | 0.38 | 0.38 | 0.39 | 0.37 | 0.40 | 0.39 | 0.37 | 0.38 | 0.37 | 0.38 | 0.39 | 0.38 | 0.38 | 0.38 | 0.01 |
| MnO | 0.13 | 0.11 | b.d.l. | b.d.l. | 0.11 | 0.12 | b.d.l. | b.d.l. | 0.12 | b.d.l. | 0.15 | 0.12 | b.d.l. | b.d.l. | 0.10 | 0.13 | b.d.l. | b.d.l. | b.d.l. | b.d.l. | 0.12 | 0.01 |
| MgO | 51.68 | 51.25 | 51.78 | 51.56 | 51.83 | 51.90 | 51.95 | 51.35 | 51.60 | 51.67 | 51.61 | 51.78 | 51.67 | 51.40 | 51.23 | 51.34 | 51.68 | 51.30 | 51.24 | 51.07 | 51.54 | 0.25 |
| Total | 99.80 | 99.64 | 100.52 | 100.00 | 100.21 | 100.44 | 100.38 | 99.85 | 100.00 | 99.97 | 99.90 | 100.71 | 100.28 | 99.86 | 99.97 | 100.01 | 100.30 | 99.71 | 100.20 | 100.07 | 100.09 | 0.28 |
| Coeficients of empirical formulae in a.p.f.u. calculated on the basis of 4 oxygens in formula unit | | | | | | | | | | | | | | | | | | | | | | |
| Si | 0.992 | 1.001 | 1.003 | 0.999 | 0.995 | 0.994 | 0.996 | 1.003 | 0.997 | 0.996 | 0.995 | 1.002 | 1.001 | 1.000 | 1.006 | 1.002 | 1.001 | 1.003 | 1.010 | 1.011 | 1.000 | 0.005 |
| Al | 0.001 | | | | | | 0.001 | | | | | | | | | | | | | | 0.001 | |
| Fe | 0.136 | 0.135 | 0.132 | 0.136 | 0.135 | 0.137 | 0.133 | 0.135 | 0.136 | 0.136 | 0.135 | 0.133 | 0.135 | 0.137 | 0.134 | 0.135 | 0.134 | 0.134 | 0.134 | 0.135 | 0.135 | 0.001 |
| Ni | 0.008 | 0.007 | 0.008 | 0.008 | 0.007 | 0.007 | 0.007 | 0.007 | 0.007 | 0.008 | 0.007 | 0.008 | 0.008 | 0.007 | 0.007 | 0.008 | 0.007 | 0.008 | 0.007 | 0.007 | 0.007 | 0.000 |
| Mn | 0.003 | 0.002 | | | 0.002 | 0.002 | | | 0.002 | | 0.003 | 0.002 | | | 0.002 | 0.003 | | | | | 0.002 | 0.000 |
| Mg | 1.869 | 1.854 | 1.854 | 1.858 | 1.865 | 1.865 | 1.865 | 1.852 | 1.861 | 1.864 | 1.864 | 1.852 | 1.856 | 1.855 | 1.845 | 1.850 | 1.856 | 1.852 | 1.839 | 1.836 | 1.856 | 0.009 |
| Total | 3.008 | 2.999 | 2.997 | 3.001 | 3.005 | 3.006 | 3.003 | 2.997 | 3.003 | 3.004 | 3.005 | 2.998 | 2.999 | 3.000 | 2.994 | 2.998 | 2.999 | 2.997 | 2.990 | 2.989 | 3.000 | 0.005 |
| Forsterite content in mol. % | | | | | | | | | | | | | | | | | | | | | | |
| Fo | 93.2 | 93.2 | 93.3 | 93.2 | 93.3 | 93.2 | 93.3 | 93.2 | 93.2 | 93.2 | 93.2 | 93.3 | 93.2 | 93.1 | 93.2 | 93.2 | 93.2 | 93.3 | 93.2 | 93.2 | 93.2 | 0.1 |

Notes: a.p.f.u. – atom per formula unit; s.d. – standard deviation; b.d.l. indicates values below detection limit; Ti, Cr, Ca, Na, and K were found to be below detection limit in all analyses.

Measurement details: microprobe CAMECA SX-100; accelerating voltage 15kV / sample current 10nA: Na$K_\alpha$ (standard: jadeite; detection limit 550 ppm), Si$K_\alpha$ (quartz; 480 ppm), Mg$K_\alpha$ (diopside; 950 ppm), K$K_\alpha$ (sanidine; 240 ppm), Ca$K_\alpha$ (diopside; 260 ppm), Ti$K_\alpha$ (rutile; 320 ppm), Al$K_\alpha$ (jadeite; 330 ppm), Mn$K_\alpha$ (rhodonite; 700 ppm), Cr$K_\alpha$ (Cr$_2$O$_3$; 600 ppm); 20kV / 10nA: Fe$K_\alpha$ (magnetite; 450 ppm), Ni$K_\alpha$ (NiSi; 230 ppm); beam diameter 2 µm.

Table 2. Conditions of sample processing. $T_1$ and $t_1$ are the temperature and time of the first heating step (in air). $T_2$ and $t_2$ are the temperature and time of the second heating step (in hydrogen). N/A – not applicable.

| Sample | $T_1$ (°C) | $t_1$ (s) | $T_2$ (°C) | $t_2$ (s) |
|---|---|---|---|---|
| Raw olivine | N/A | N/A | N/A | N/A |
| 400h1 | 400 | 3600 | 500 | 3600 |
| 450h1 | 450 | 3600 | 500 | 3600 |
| 500h1 | 500 | 3600 | 500 | 3600 |
| 550h1 | 550 | 3600 | 500 | 3600 |
| 600h1 | 600 | 3600 | 500 | 3600 |
| 650h1 | 650 | 3600 | 500 | 3600 |
| 700h1 | 700 | 3600 | 500 | 3600 |
| 750h1 | 750 | 3600 | 500 | 3600 |
| 850s30 | 850 | 30 | 500 | 3600 |

Table 3. Overview of sample spectral parameters, saturation magnetization ($J_s$) and calculated npFe$^0$ concentration. Iron concentration is calculated as the sample saturation magnetization divided by the saturation magnetization of pure iron (218 Am$^2$/kg). * – not possible to reliably determine.

| Sample | Slope (µm$^{-1}$) | 1µm depth | Albedo at 550 nm | 1um band center (nm) | 1µm band width at half depth (nm) | $J_s$ (mAm$^2$/kg) | Fe (wt%) |
|---|---|---|---|---|---|---|---|
| Olivine | 0.031 | 0.21 | 0.78 | 1053 | 466 | 0 | 0.0 |
| 400h1 | 0.15 | 0.18 | 0.61 | 1048 | 356 | 16.44 | 0.0075 |
| 450h1 | 0.17 | 0.18 | 0.60 | 1048 | 347 | 18.43 | 0.0085 |
| 500h1 | 0.25 | 0.20 | 0.42 | 1048 | 306 | 24.59 | 0.011 |
| 550h1 | 0.22 | 0.15 | 0.52 | 1046 | 265 | 27.78 | 0.013 |
| 600h1 | 0.36 | 0.16 | 0.34 | 1045 | 222 | 31.89 | 0.015 |
| 650h1 | 0.41 | 0.10 | 0.28 | 1041 | 165 | 49.71 | 0.023 |
| 700h1 | 0.57 | 0.026 | 0.16 | 1031 | 118 | 128.7 | 0.059 |
| 750h1 | 0.69 | 0.00 | 0.11 | 970 | * | 253.2 | 0.12 |
| 850s30 | 0.36 | 0.034 | 0.28 | 1034 | 139 | 106.3 | 0.049 |

Table 4. Modified Gaussian model parameters of the olivine absorption band. C1-3 are the three Gaussian centers. FWHM1-3 are the full widths at half maximum. S1-3 are the Gaussian strengths. RMSD is the root mean square deviation.

| | C1 (nm) | C2 (nm) | C3 (nm) | FWHM1 (nm) | FWHM2 (nm) | FWHM3 (nm) | S1 | S2 | S3 | RMSD |
|---|---|---|---|---|---|---|---|---|---|---|
| Raw olivine | 826 | 1022 | 1229 | 124 | 221 | 465 | -0.0863 | -0.239 | -0.276 | 2.27E-02 |
| 400h1 | 837 | 1026 | 1202 | 170 | 181 | 436 | -0.093 | -0.195 | -0.265 | 4.73E-03 |
| 450h1 | 834 | 1025 | 1201 | 166 | 184 | 450 | -0.0916 | -0.192 | -0.27 | 7.28E-03 |
| 500h1 | 847 | 1032 | 1194 | 214 | 169 | 442 | -0.126 | -0.188 | -0.303 | 3.77E-03 |
| 550h1 | 835 | 1028 | 1195 | 175 | 180 | 454 | -0.0821 | -0.161 | -0.237 | 5.88E-03 |
| 600h1 | 853 | 1039 | 1187 | 265 | 160 | 464 | -0.104 | -0.134 | -0.277 | 2.86E-03 |
| 650h1 | 846 | 1040 | 1171 | 280 | 154 | 487 | -0.0697 | -0.101 | -0.226 | 2.77E-03 |

Fig. 1. The relation between the temperature of first heating step (in air) and amount of iron produced in form of iron nanoparticles. The heating duration was 1 hour for all displayed samples.

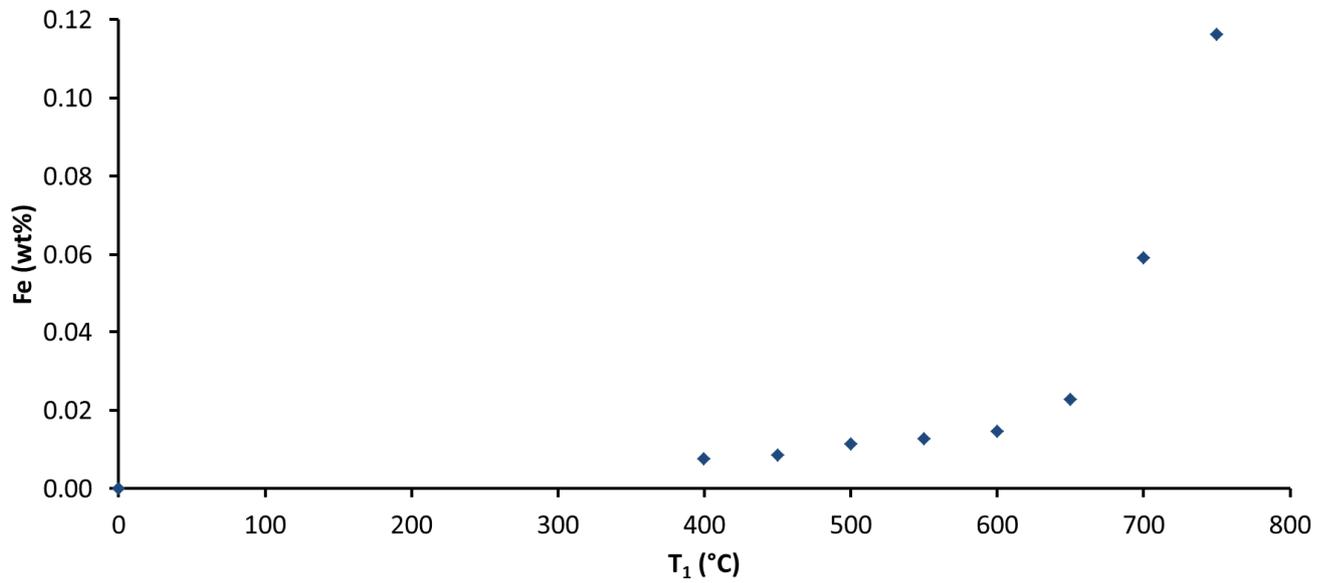

Fig. 2. TEM image of ~5-20 nm sized nanoparticles on olivine powder grains (sample 600h1).

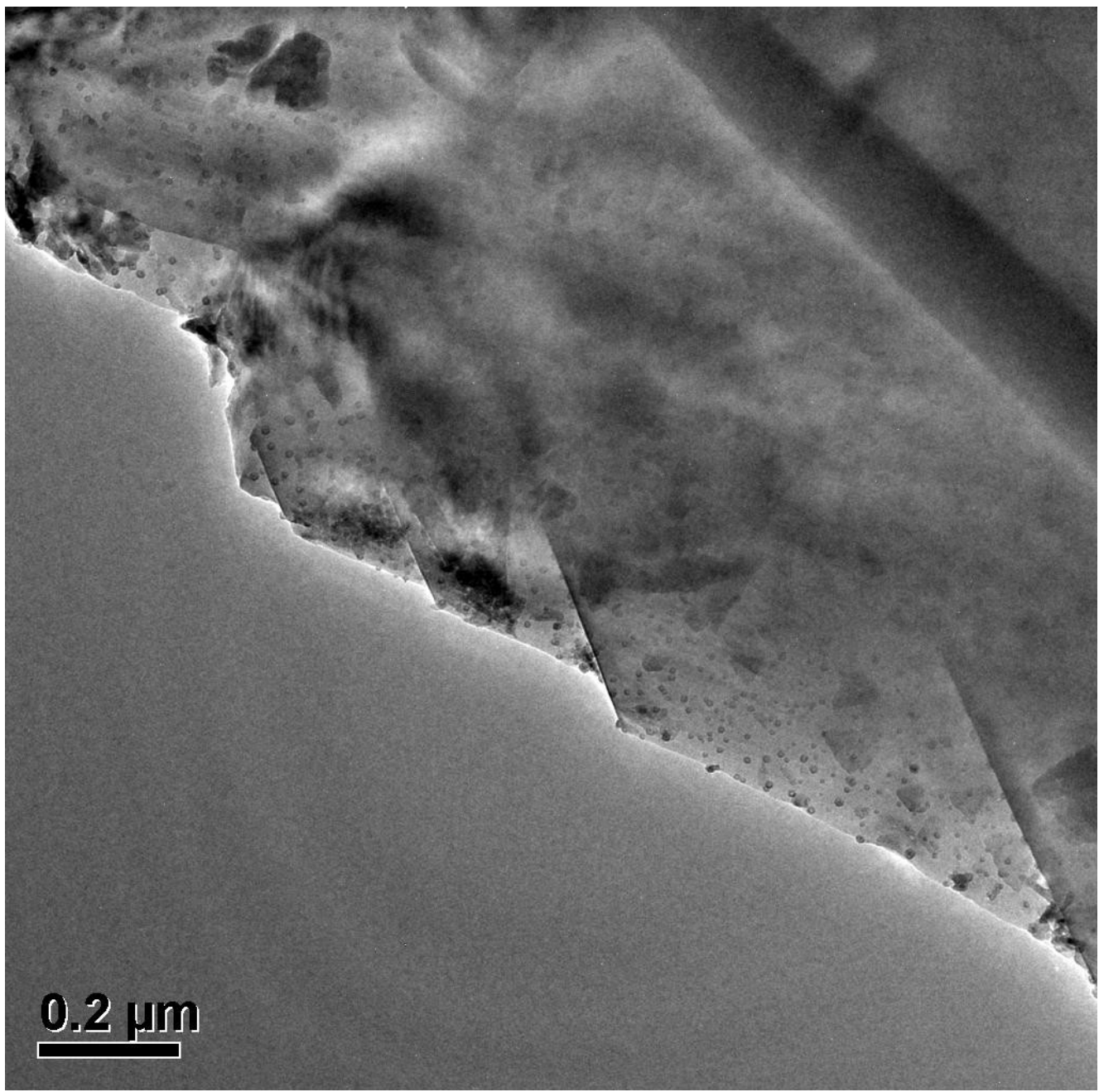

Fig. 3. TEM image of nanoparticles on olivine powder grains of the 850s30 sample. Two populations can be observed (~5-20 nm and ~40-50 nm).

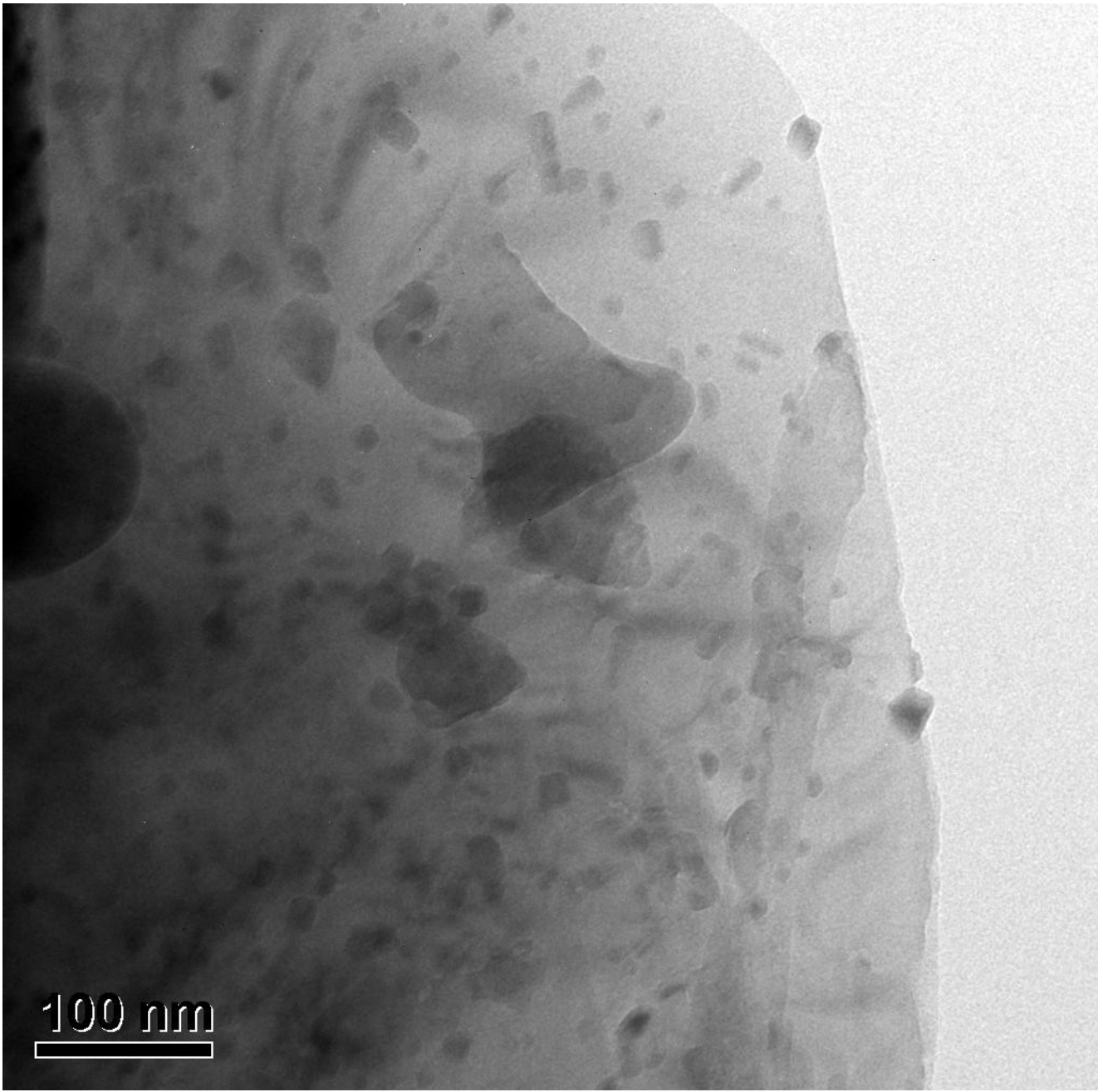

Fig. 4. High resolution TEM image of a single iron nanoparticle. The lattice fringes are highlighted by yellow circle. The lattice spacing is 0.19 ±0.05 nm.

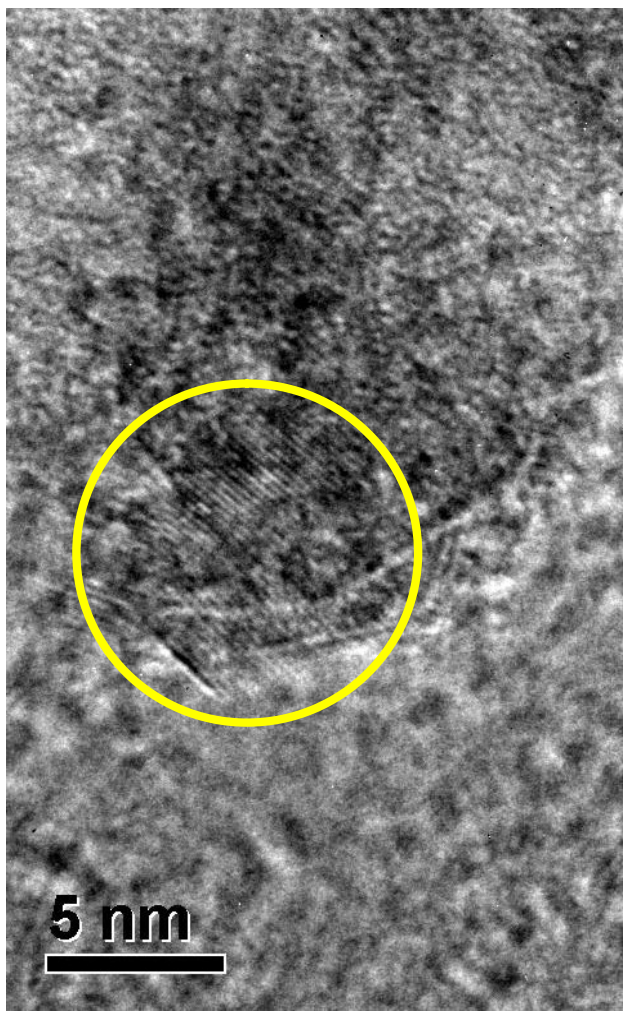

Fig. 5. STEM EDX spectra of an individual iron nanoparticle. These data show that iron nanoparticle oxidation was kept to minimum since only a minor peak of oxygen is observed and that is most likely caused by the olivine background (also indicated by a minor peak of silica and magnesium). Peak of copper is the signal of the sample holder.

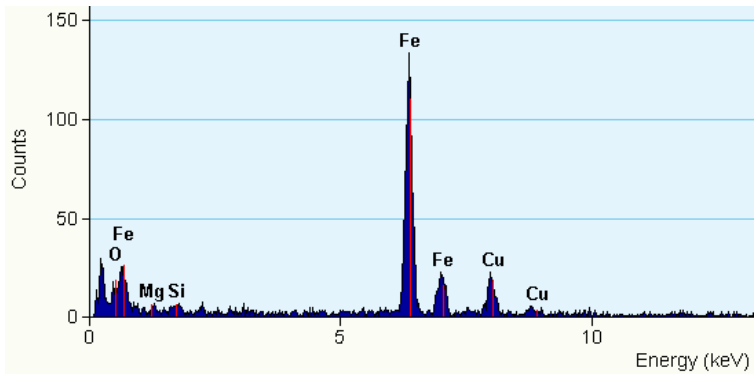

Fig. 6. VIS-NIR reflectance spectra (up – absolute values, down – normalized at 550 nm) of the fresh and modified olivine samples with increasing npFe$^0$ concentration (in wt%) as determined from the saturation magnetization. All samples show the continuous trend of 1 μm olivine absorption band reduction and increasing red slope with increasing amount of the npFe$^0$. The 850°C sample does not follow the increasing red slope trend because it contains additional larger npFe$^0$ particles that do not contribute to the red slope.

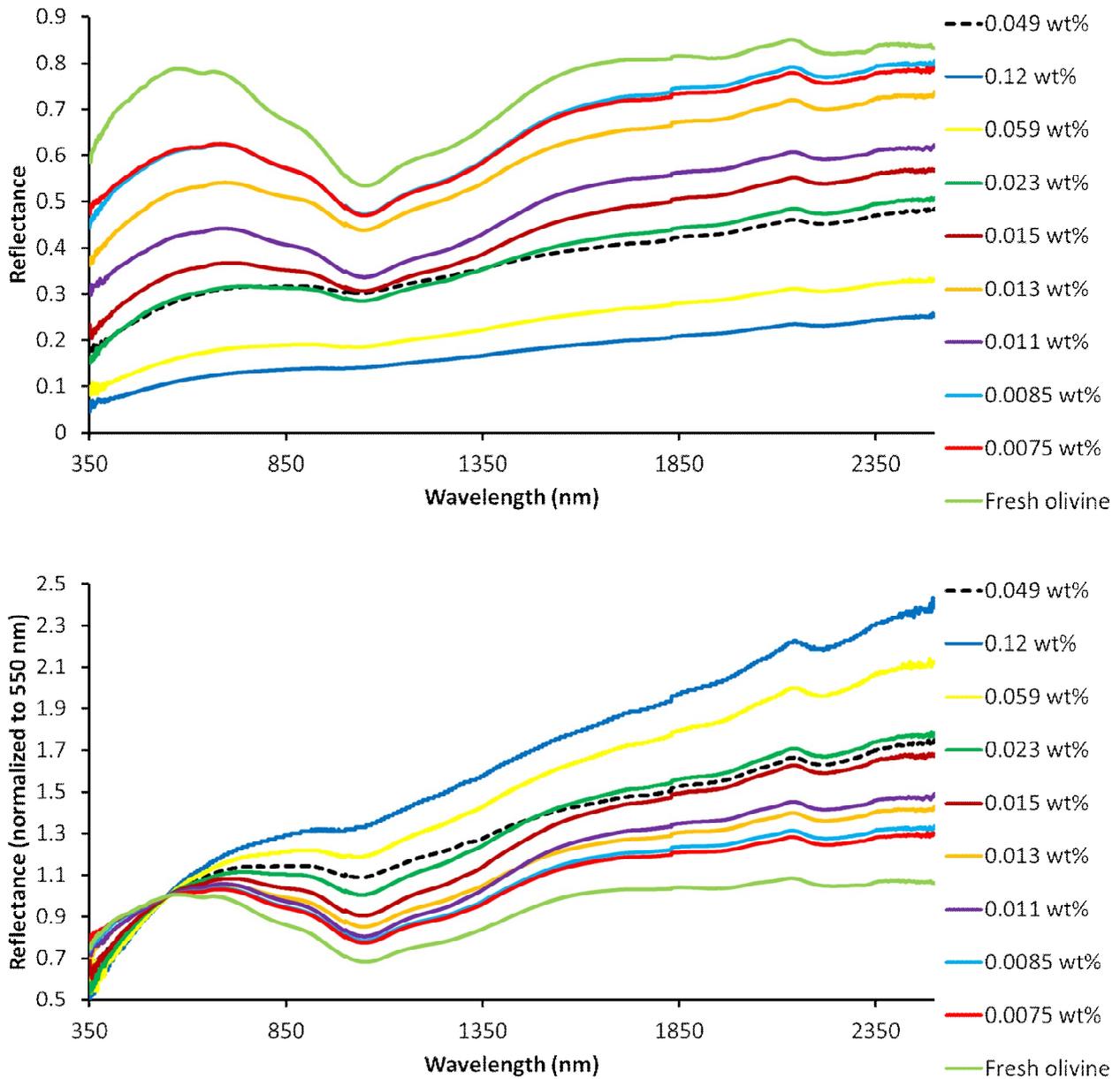

Fig. 7. The linear trend (black line) between the npFe$^0$ amount and the position 1 μm band center. $R^2$ is the root mean square deviation of the fit.

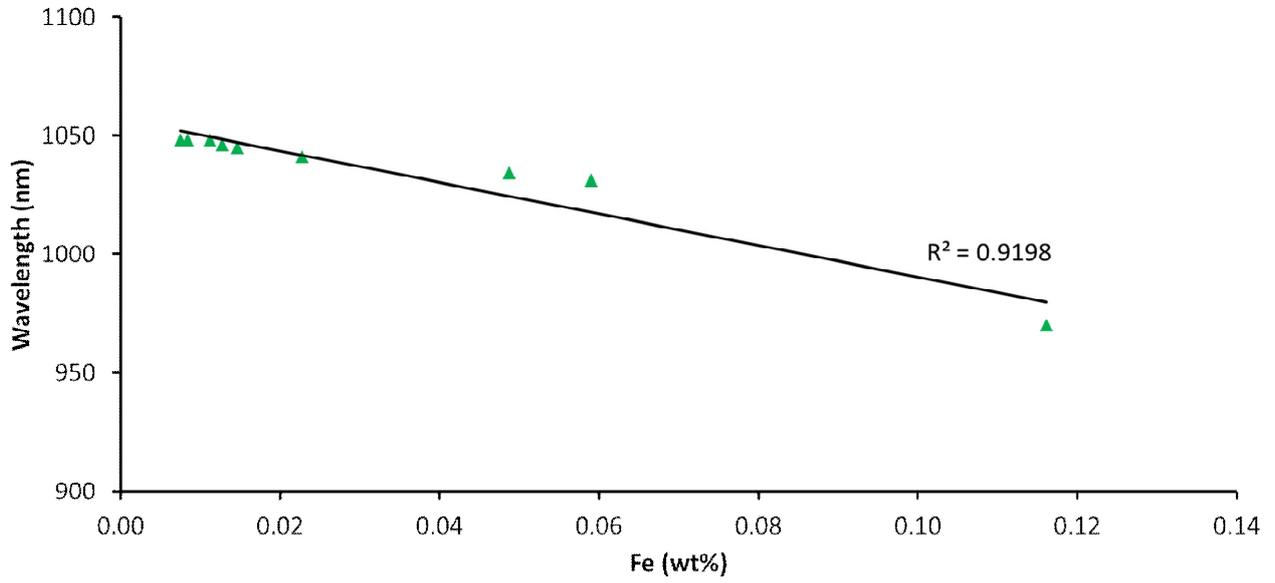

Fig. 8. The logarithmic trend (black lines) between the npFe$^0$ amount and the 1 μm band depth and the reflectance at 550 nm. $R^2$ is the root mean square deviation of the fit.

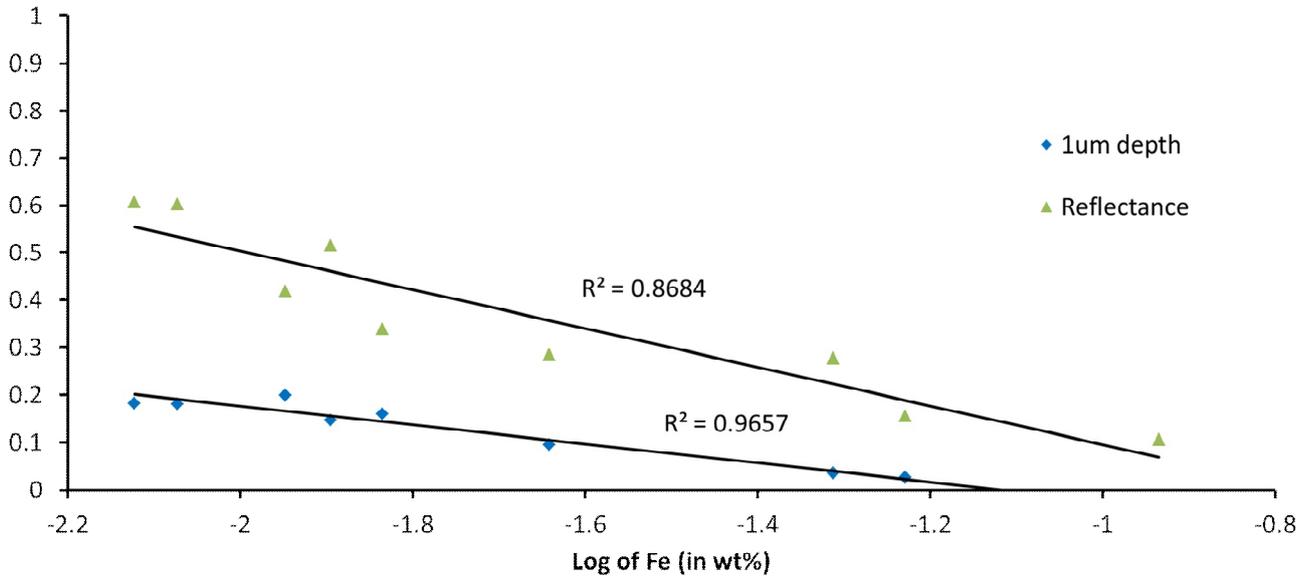

Fig. 9. The logarithmic trend (black line) between the npFe$^0$ amount and the spectral slope. The 850s30 sample (highlighted by a yellow circle) with additional population of larger npFe$^0$ particles does not fully follow the reddening trend. R$^2$ is the root mean square deviation of the fit.

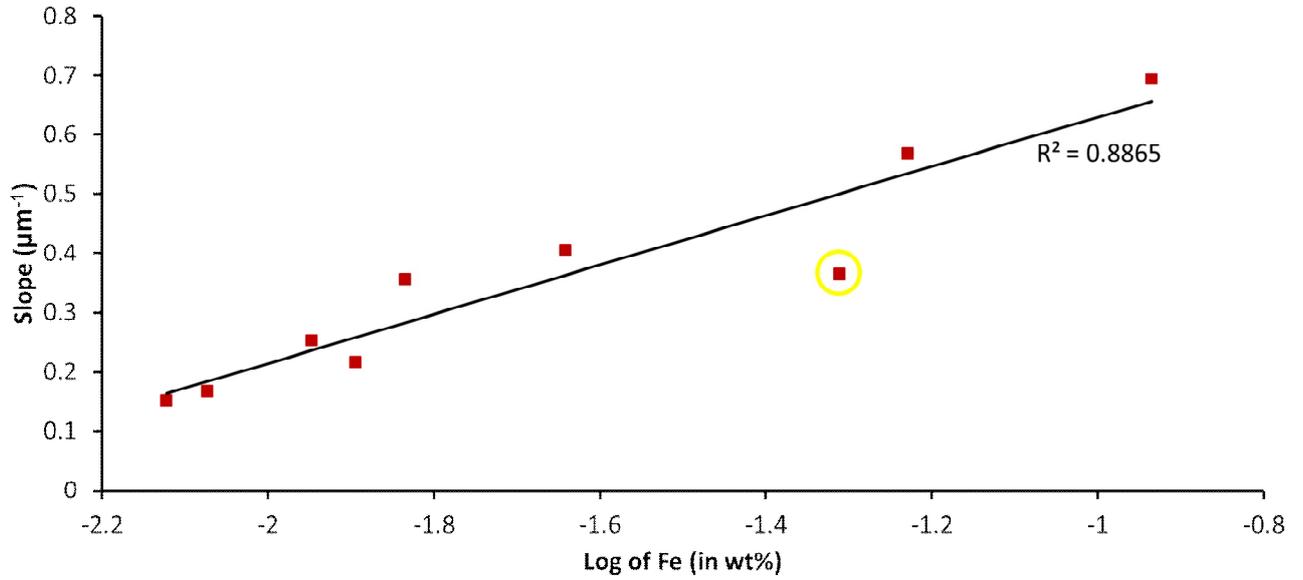

Fig. 10. The logarithmic trend (black line) between the npFe$^0$ amount and the 1 µm band width at half depth. R2 is the root mean square deviation of the fit.

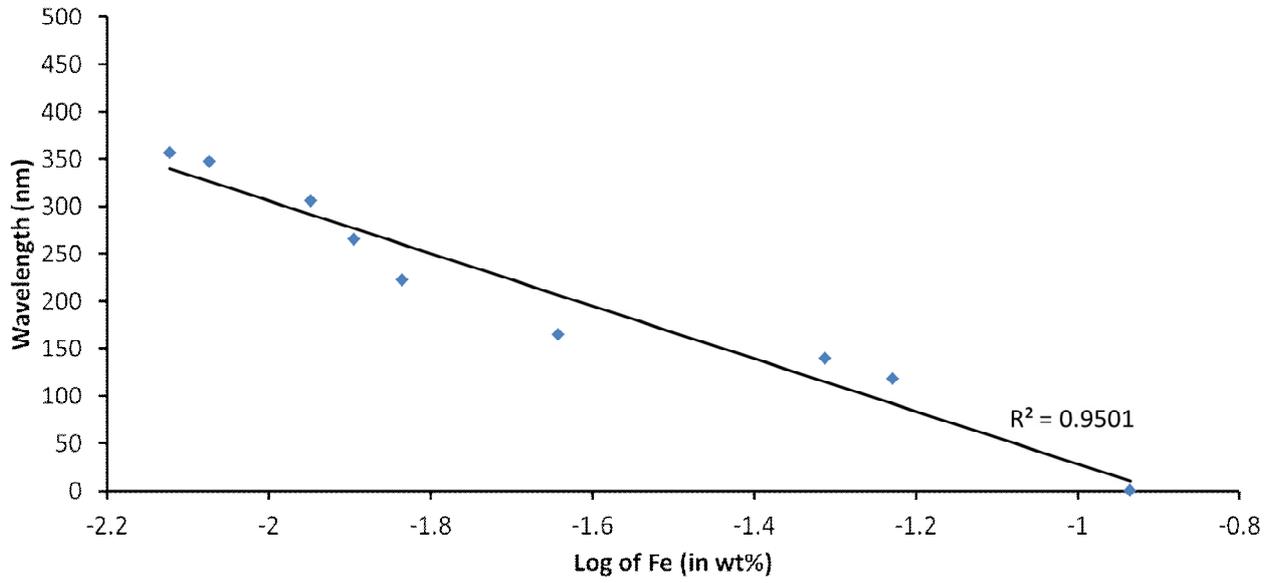

Fig. 11. Positions of the Gaussian band centres (C1-3) as a function of increasing npFe[0] content.

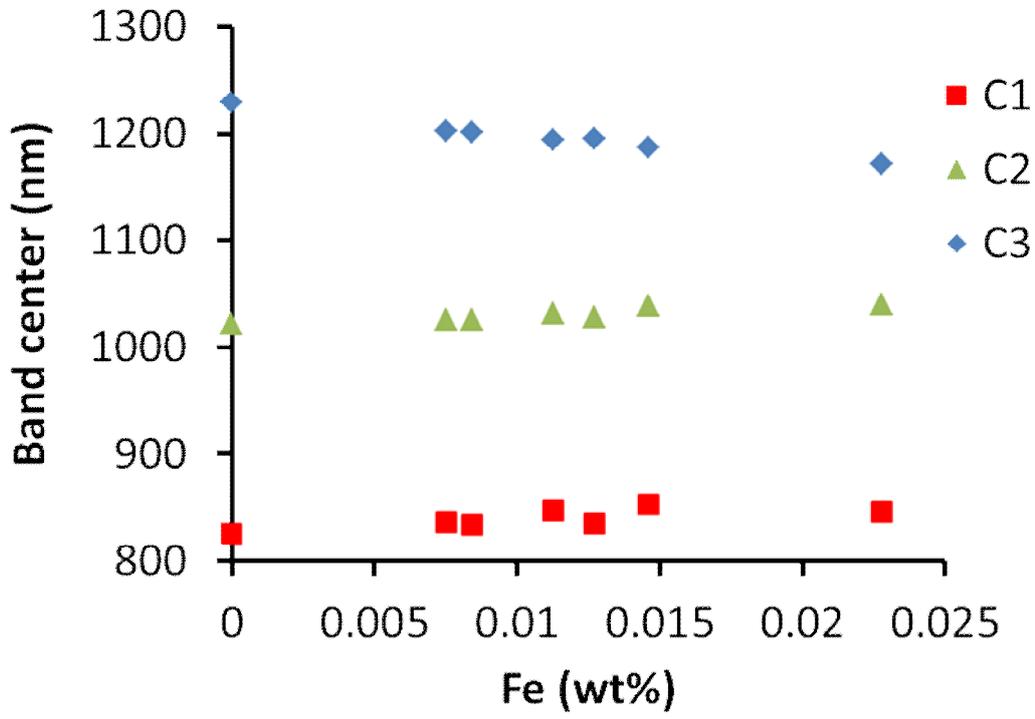

Supplementary material

Modified Gaussian model figures for samples listed in Table 4. Orange – measured spectrum, black – modeled spectrum, blue – gaussian bands, red – continuum, pink – residual error spectrum.

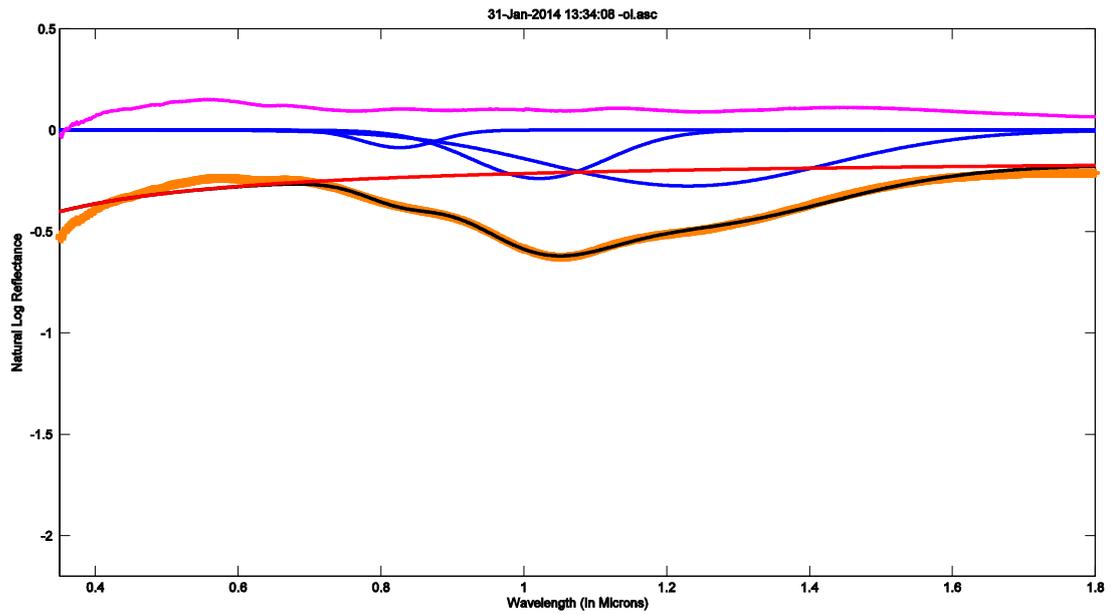

Fig. S1. Fresh olivine.

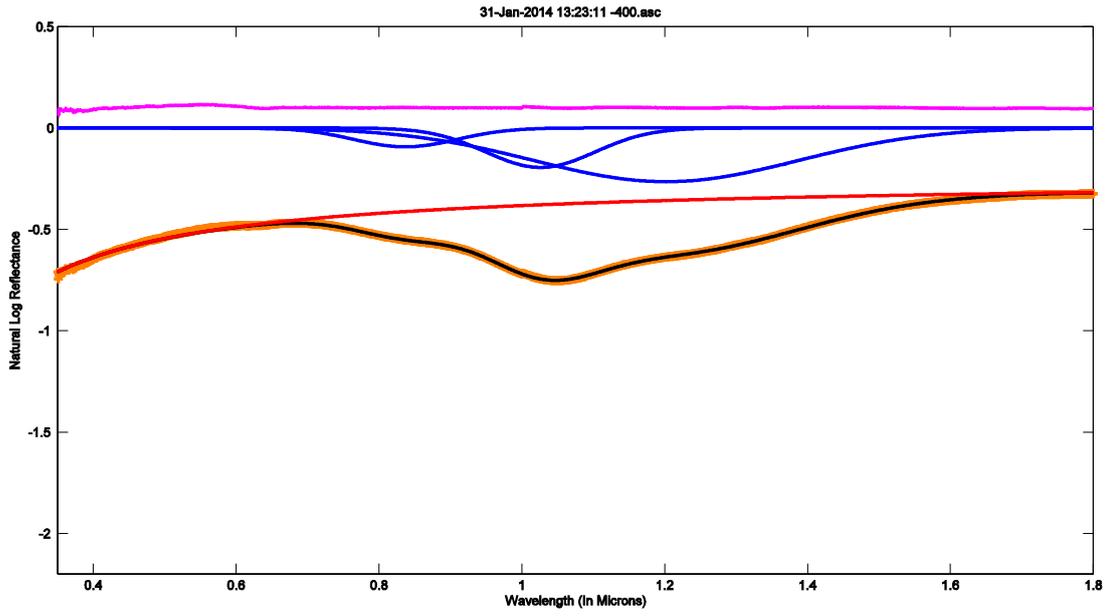

Fig. S2. Sample 400h1.

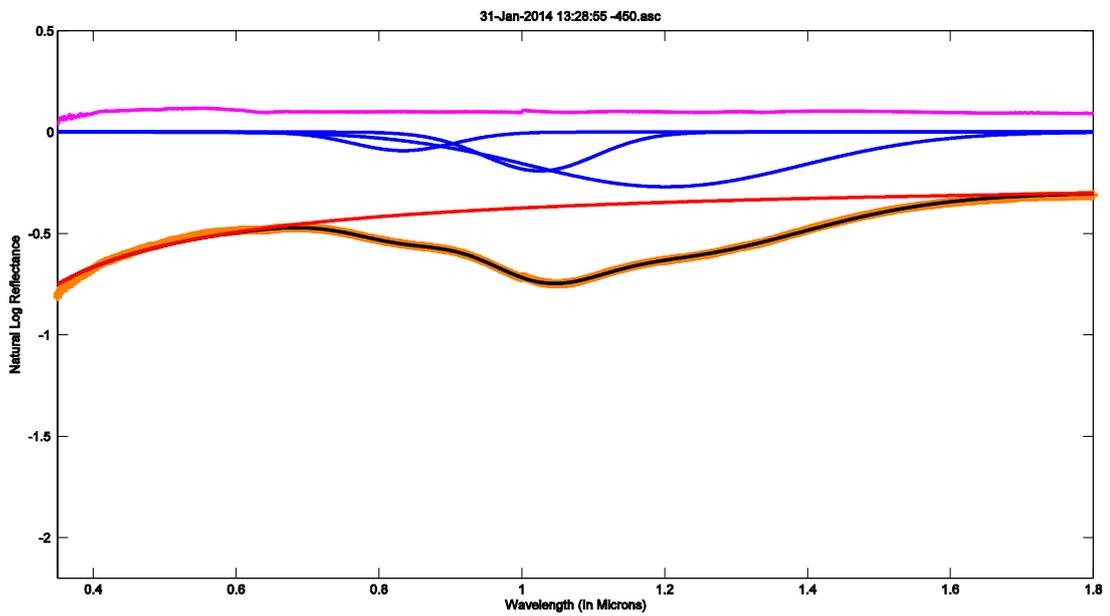

Fig. S3. Sample 450h1.

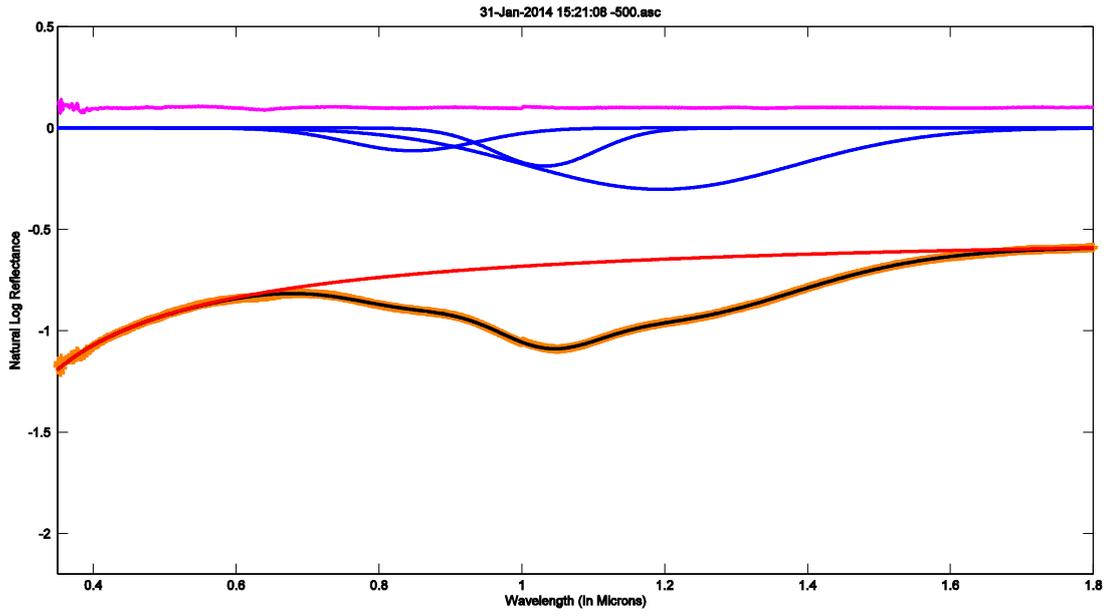

Fig. S4. Sample 500h1.

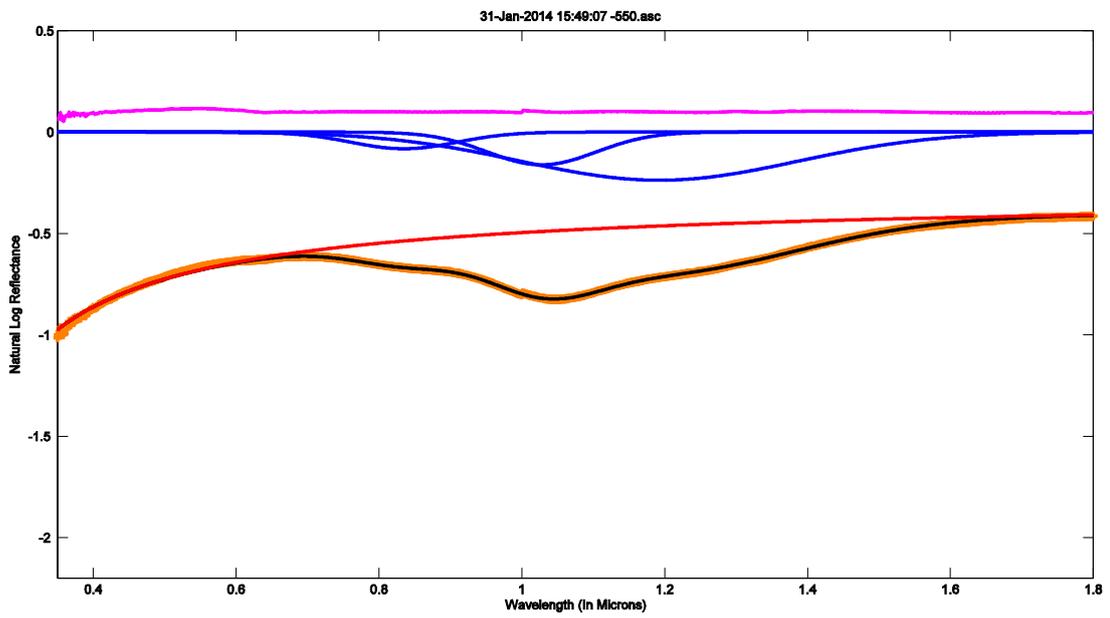

Fig. S5. Sample 550h1.

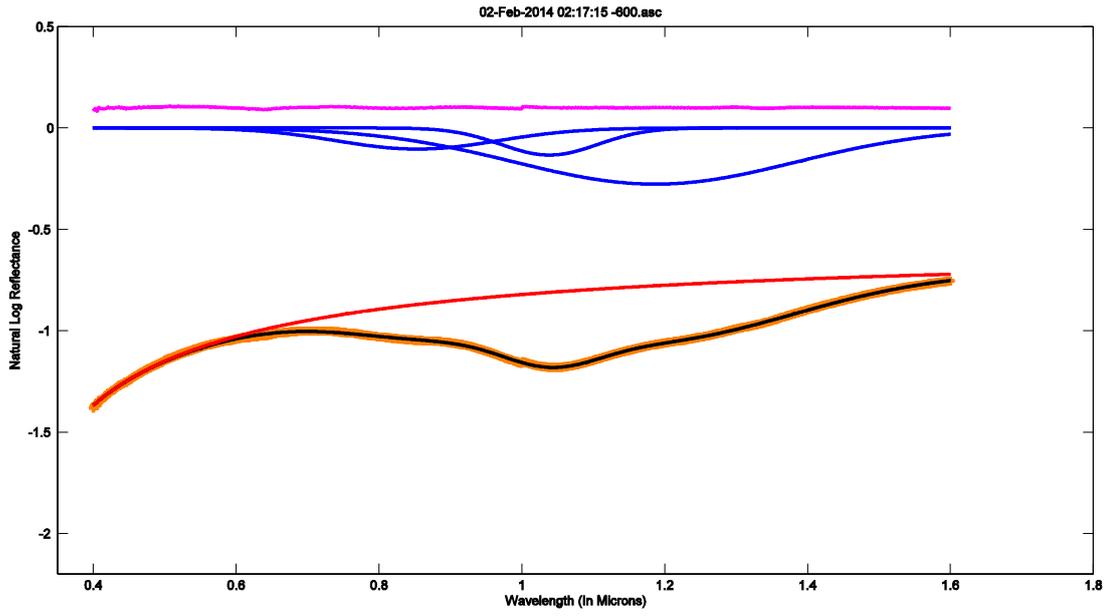

Fig. S6. Sample 600h1.

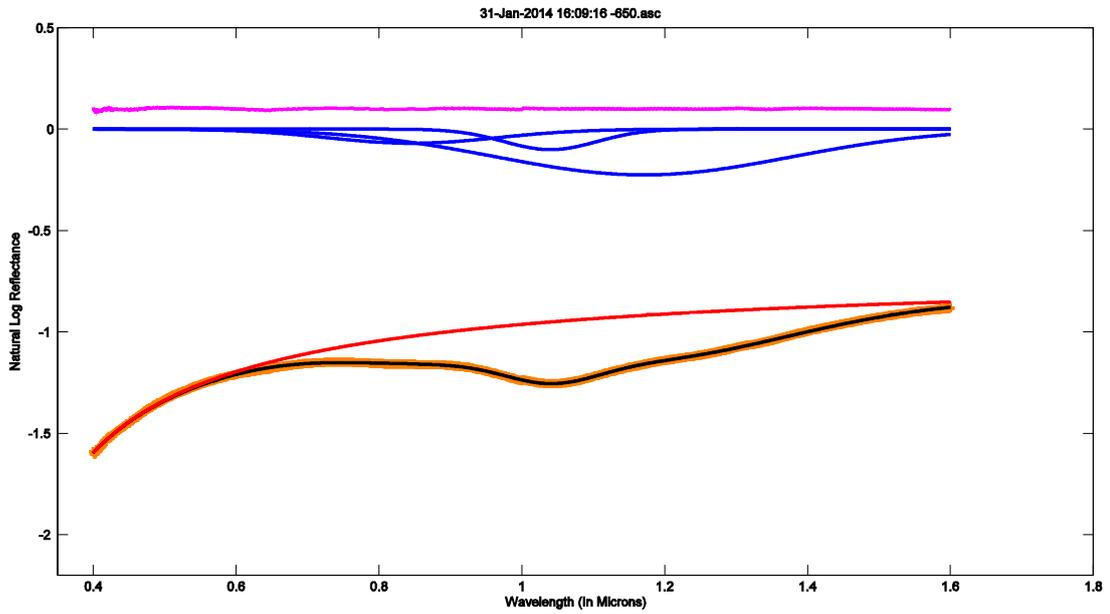

Fig. S7. Sample 650h1.